\documentclass[10pt, conference, compsocconf]{IEEEtran}
\usepackage[sort&compress,square,numbers]{natbib}
\usepackage[hidelinks]{hyperref} 

\begin{document}

\title{Towards a Security Engineering Process Model for Electronic Business Processes}

\author{\IEEEauthorblockN{J\"orn Eichler}
\IEEEauthorblockA{Fraunhofer Institute for Secure Information Technology (SIT)\\
D-64295 Darmstadt, Germany\\
Email: joern.eichler@sit.fraunhofer.de}}

\maketitle


\begin{abstract}
Business process management (BPM) and accompanying systems aim at enabling enterprises to become adaptive. 
 In spite of the dependency of enterprises on secure business processes, BPM languages and techniques provide only little support for security. 
Several complementary approaches have been proposed for security in the domain of BPM. Nevertheless, support for a systematic procedure for the development of secure electronic business processes is still missing. In this paper, we pinpoint the need for a security engineering process model in the domain of BPM and identify key requirements for such process model.
\end{abstract}

\begin{IEEEkeywords}
security engineering; process model; requirements, business process management
\end{IEEEkeywords}

\IEEEpeerreviewmaketitle

\section{Introduction}

Business processes are the way organizations do their work -- a set of activities carried out to accomplish a defined objective. Therefore, the design, administration, enactment, and analysis of business processes -- subsumed under the term business process management (BPM) -- are vital challenges to organizations. Business process management systems (BPMS) are seen as important facilitators for the necessary alignment of people and organizational resources. BPMSs enable organizations to become adaptive enterprises: They allow for a faster reaction to environmental and market changes and support proactive innovation of products and services.
\cite{davenport2005coming,tallon2008inside}

Consequently, BPM supported by information systems has seen an ongoing development in the last decades. With respect to modeling languages and techniques, a multitude of approaches has been introduced. At the same time, BPMS developed from simple information systems to capture and administrate process models to feature-rich BPM suites that support also simulation, execution, and controlling of business process instances. Hence, the ability for organizations to manage business processes is well supported by today's software industry. \cite{recker2009business}

Business processes are closely connected with assets of the respective organization. Observation, manipulation, and disruption of business processes might threaten these assets or even the existence of the organization itself. Thus, security of business processes ought to be of high importance for every organization. But in spite of this dependence on secure business processes, BPM languages and techniques provide only little support to express security needs or controls applied.

\section{Security and Electronic Business Processes}

In reaction to this need, several approaches have been developed to support security in the context of BPM. Most approaches address one of two issues: the analysis of security (and safety) properties of business process models or the specification of security requirements and controls for electronic business processes \cite{atluri2008security}.

By contrast, the support for a systematic procedure to develop secure electronic business processes is weak. The few existing approaches do not address actual runtime environments and enforceable controls \cite{roehrig2004security,neubauer2010workshop}, support only specific  activities like security requirements engineering \cite{herrmann2006security,rodriguez2011secure}, or do not provide any guidance for their application \cite{mana2003business}. Also general approaches applied to electronic business processes display similar issues \cite{juerjens2009security}.

This situation might be attributed to security engineering in general: As a discipline -- commonly defined as \textit{``building systems to remain dependable in the face of malice, error, or mischance'' } \cite{anderson2001security} -- it is considered to be still in its infancy. At present, mostly top-down approaches from the software engineering domain are adopted and enhanced with security-specific technologies and methods.

With regard to BPM, lack of support for security engineering is endangering one of its main objectives: allowing enterprises to react faster and to continuously innovate products and services. Currently, enterprises either have to choose to focus on the protection of their assets or to develop and deploy their electronic business processes with little security expertise and support by applicable methods. The first option requires security professionals to secure the electronic business processes individually and manually which implies investments in terms of time and money and threatens the adaptability gained with the application of BPMS. The second option exposes the enterprises' assets to malice and mischance. Industrial experiences from our Security Test Lab as well as academic studies analyzing industrial security engineering practices indicate that enterprises tend to choose the latter option \cite{vaughn2002empirical}.

\section{Requirements for a Security Engineering Process Model Addressing BPM}

As a first step to bridge the gap between (executable) business process models and secure electronic business processes, we provide a set of requirements for a security engineering process model in the domain of BPM. These requirements stem from fundamental ideas of BPM: separation of technical and domain aspects (allowing domain experts to work largely independently from developers), independence from development methodologies, and applicability notwithstanding environmental heterogeneity. 

Key issues for a general security engineering process model are separation of requirements and controls, traceability, correctness, completeness, and iterative applicability on different levels of abstraction. Core activities encompass security requirements elicitation, threat modeling and evaluation, control design, and validation. \cite{breu2003key}

To align a security engineering process model with BPM we identify the following requirements:
\begin{enumerate}
\item Separation of (initial)  activities for security professionals and (recurring) activities for security nonprofessionals 
\item Consistent coverage of all activities with detailed guidance
\item Utilization of models and adequate tooling to separate the security analysis from design and implementation
\item Possibility to integrate the security engineering process model with different development approaches
\end{enumerate}

We envision a security engineering process model that aids security professionals to prepare an environment for domain experts, providing common threats, evaluation criteria, and their countermeasures supported by business process engines. The security engineering process model supports domain experts to identify and evaluate security requirements utilizing business process models as primary input, to select from a restricted set of applicable controls, and to configure the business process engines correspondingly.

\section{Conclusion}

BPMS enable enterprises to become adaptive and are well supported by today's software industry. Although an active research community proposed several approaches to address the need for security in the domain of BPM, support for a systematic procedure for the development of secure electronic business processes is still missing. We identify four key requirements for a security engineering process model that is able to bridge the gap between (executable) business process models and secure electronic business processes. Currently we are working on such a process model as well as supporting modeling languages and tooling.

\section*{Acknowledgment}

The work presented was developed in the context of the project Innovative Services for the Internet of the Future (InDiNet, ID 01IC10S04F) which is funded by the German Federal Ministry of Education and Research.

\IEEEtriggeratref{8}

\bibliographystyle{IEEEtran}
\bibliography{FA_Eichler_FhI-SIT}

\end{document}